\documentclass[prl,a4paper,onecolumn,preprint,footinbib,showpacs]{revtex4}
\usepackage{amsmath,bm}
\newcommand{\qo}[1]{\textquotedblleft #1\textquotedblright}
\begin{document}
\title{Resolution of the Cosmological Constant Paradox}
\author{Francesco De Martini}
\affiliation{Dipartimento di Fisica, Universita' la Sapienza, I-00185 Roma, Italy} 
\affiliation{Accademia dei Lincei, via della Lungara 10, I-00165 Roma, Italy}
\email{francesco.demartini@uniroma1.it}
%%%

%%%%%%%%%%%%%%%%%%%%%%%%%%%%%%%%%%%%%%%%%%%%
\date{version: 23 February, 2018}
%%%%%%%%%%%%%%%%%%%%%%%%%%%%%%%%%%%%%%%%%%%%

\begin{abstract}
The standard electroweak theory of leptons and the conformal groups of spacetime Weyl's transformations are at the core of a general-relativistic, conformally-covariant scalar-tensor theory aimed at the resolution of the most intriguing enigma of modern Physics: the "cosmological constant paradox" (hereafter: $"\Lambda$-paradox"). A Higgs mechanism within a spontaneous symmetry breaking process offers formal connections, via an "effective potential" $V_{eff}$, between some relevant  properties of the elementary particles and the dark energy content of the Universe.  The non-integrable application of the Weyl's geometry leads to a Proca equation accounting for the dynamics of a vector-meson proposed as an an optimum candidate for Dark Matter.  The resolution of the paradox is found for all exponential inflationary potentials and is consistent with the recent experimental data of the PLANCK Mission.  The average vacuum-energy density in the Universe is found: $\rho_{vac}=(3.44\cdot10^{-3})^4 (eV)^{4}$ and the value of the "cosmological constant": $\Lambda={3.86}\cdot{10^{-64}}{ (eV/c^{2})^{2}}$. The result of the theory: $\Lambda=6|V_{eff}|$ shows that the paradox is determined by the  algebraic "mismathch" between two large counteracting functions of the "inflaton" field contributing to $V_{eff}$.  The critical stability of the Universe is discussed.\\


\end{abstract}
%%3}
\maketitle
%% Letter text
\section{1.
Introduction}
A huge step forward in theoretical cosmology, and a landmark of modern science, was the proposal by A. Starobinsky followed by A.Guth, A. D. Linde, A. Albrecht and P.J. Steinhard~\cite{Starobinsky1980, Guth1981,Linde1982,AlbrechtSteinhardt1982} of the \textit{inflation}, an epoch of fast accelerating expansion of the early Universe that caused the Universe to expand through about 70 $e$-folds in a small fraction of a second. 
This expansion, driven by a scalar field called \qo{inflaton}, was originally argued to solve the problem of why the universe is so smooth at large scales.  Over the years, the undeniable success of this idea was somewhat questioned by the failure of finding the physical mechanism underlying the nature of the {\textit{inflaton}} concept~\cite{Dodelson2003,Weinberg1972}. In a recent letter we claimed that the fundamental nature of this scalar field is indeed \textit{geometrical}, based on the conformal differential geometry introduced by Hermann Weyl~\cite{Weyl1952,Lord1979,De Martini2016,foot:Weyl}.\\
 In the present work we analyze these concepts and apply them to the resolution of a disturbing enigma of modern Science: the "cosmological constant paradox" (hereafter:$"\Lambda$-paradox"). This puzzle, well known since many decades and still lacking of a consistent resolution, consists of the  humongous discrepancy existing between the value of the energy density of the vacuum-field in the Universe evaluated by the methods of quantum field theory (QFT) and the measured value of this same quantity. The resolution of this enigma is particularly needed at the present times where the "quantum gravity" endeavour is at the top of the concern of a large class of physicists. The paper is organized as follows: In Section 2 some aspects of the Weyl geometry are analyzed in the framework of a quantum dynamical problem involving the interaction of the "inflaton" and of the Higgs field. In particular the definition of an "effective potential" driving the inflaton field is introduced. In Section 3  the QFT of leptons is outlined in order to analyze the symmetry-breaking process within the dynamics of the vacuum-field in the Universe. Section 4 is devoted to the analysis of the $\Lambda$-paradox and there a consistent resolution is proposed. In Section 5 the solution is analyzed in the perspective of recent experimental measurements by the PLANCK Mission. There the critical instability of the Universe are discussed. In Section 6 a mathematical extension of the Weyl geometry suggests a possible application of the theory to the cosmological "Dark Matter".\\ \\  

%%%%%%%
%%%%%%%%%%%%%%%%%%%%%%%%%%%%%%%
\section{2.
The Weyl geometry and the cosmological inflation}

 The Weyl geometry rests on a statement that may be cast in the simple form:  \qo{\textit{All Laws of Physics are conformally-covariant}} (or \textit{Weyl covariant} or, in short, \textit{co-covariant}).\\ Our conjecture regarding the physics-geometry connection is as follows. According to Weyl, the parallel displacement from two infinitely nearby points $P$ and $P + dP$ in spacetime acts on the length $\ell$ of any vector by inducing a "calibration" (i.e. "gauge") change  $\delta\ell=-\ell\phi_\rho dx^\rho$, where $\phi_\rho$ is a universal \qo{\textit{Weyl vector}}, defined in the whole spacetime. Indeed the structure of the Weyl geometry implies two forms: the \textit{quadratic} Riemannian one, $g_{\rho\sigma}dx^\rho dx^\sigma$, being $g_{\rho\sigma}=g_{\sigma\rho}$ the metric tensor, and the Weyl's \textit{linear} one $\phi_\rho dx^\rho$, which is nonexistent in Riemann's geometry. The parallel displacement is \textit{integrable} iff a scalar \qo{\textit{Weyl potential}} $\phi$ exists such as: $\phi_\rho=\partial_\rho\phi$.  The {\textit{Weyl vector}} $\phi_\rho(x)$ and the {\textit{Weyl potential}} $\phi(x)$, defined in the 4-D, $(x^\rho)$ space-time play a basic role in the present work since there the inflaton $\underline{\textit{is identified}}$ with $\phi(x)$. The inflationary physics is identified with geometry.  The field  $\phi$ is a gauge-field since the Weyl geometry consists of an abelian, local, scale-invariance gauge theory implying the group of conformal transformations: \{$({\phi_\rho} \rightarrow{\phi_\rho + \partial_\rho\lambda(x)})$;
 $(g_{\rho\sigma}\rightarrow{ e^{2\lambda(x)}g_{\rho\sigma}})$\}~\cite{Weyl1952,Quigg1983}. In this way the rigid scale-invariance that requires the conformal flatness due to the Lorentz-invariance is promoted to a \textit{local} scale-invariance in order to get the Poincare'-invariant field theory in \textit{arbitrary} curved spacetimes ~\cite{Iorio1997,Dengiz2017}.  Details on the Weyl geometry are given in:~\cite{Weyl1952,De Martini2016,francesco2017,fran2018,foot:Weyl,Scholz2012}. \\
The self-interaction of the inflaton field $\phi$ is driven by a potential function $V_\Lambda(\phi)$ which is determined by the quantum vacua of all fields in the Universe ~\cite{Amendola2010}. The function $V_\Lambda(\phi)$ brings within the "classical" structure of General Relativity (GR) the emblematic "quantum-mechanical vacuum", or "zero-point" concept. As suggested by Steven Weinberg we impose in the theory the general stability condition for the $\phi-field$ ~\cite{Weinberg1972}:
\begin{equation}\label{eq:Beltrami}
\nabla_B\phi = - V'_\Lambda(\phi)
\end{equation}
being $\nabla_B$ the Laplace-Beltrami differential operator and: $V'_\Lambda(\phi)\equiv{\frac{\partial V_\Lambda(\phi)}{\partial\phi}}$ ~\cite{foot:Weyl}. The dimensionless scalar quantities $\phi$ and $V_\Lambda$ appearing in Eq.~(\ref{eq:Beltrami}) must now be rescaled to assume the correct dimensionality: $\phi\rightarrow{K\phi}$ and: $V_\Lambda\rightarrow{K^2V_\Lambda}$ with:
 $ K^2= \frac{8\pi{G_N}}{c^4}$ being:  $G_N = 6.708{\cdot}10^{-39}\times [\hbar{c}] (GeV/c^2)^{-2}$  the gravitational constant ~\cite{De Martini2016}. In this way in the following theory the rescaled $\phi$ will assume the dimensionality of a "mass", i.e.: $[\phi]=[L^{-1}]$ and $V_\Lambda$ the dimensionality of a $(mass)^2$, i.e.: $[V_\Lambda]=[L^{-2}]$. \\\

All these concepts will be applied to a very general Weyl-Dirac conformal theory involving the mass $m_E$ of an elementary particle, e.g. an electron~\cite{Dirac1973}. 
The mass of the particle  is expressed in form of a "mass field" $m_E\rightarrow{k_E\cdot\mu(x)}$, being the dimensionless coupling constant $k_E$ an intrinsic particle's property, and $\mu(x)$  a real scalar field, function of space-time, with weight $W(\mu)=-1$ ~\cite{Jammer2000}. We define a dimensionless constant: $({\alpha})=[(c\sqrt{2})^5\cdot{G_F} \hbar]\times[8{\pi}{G_N}]^{-1}$ where: $ G_F = 1.166{\cdot}10^{-5}  (GeV)^{-2}$ is the Fermi constant~\cite{Quigg1983}. By applying the Dirac's Lagrange-multiplier method, the most general scalar-tensor form of the co-covariant Lagrangian density in $D=4$ can be expressed as~\cite{Dirac1973,Lord1979}:
\begin{equation}\label{eq:LL}
\hat{L} = \sqrt{-g}\left\{\alpha\mu^2\left[\bar R +\eta(n) V_\Lambda(\phi)\right]-{D_{\rho}\mu}{D^{\rho}\mu}+{|\tilde{\lambda}}|\mu^4-\frac{1}{4}\beta^2\phi_{\rho\sigma}\phi^{\rho\sigma}-\frac{1}{4}f^{\rho\sigma}f_{\rho\sigma}-\frac{1}{4}F^{l}_{\rho\sigma}F^{l\rho\sigma}\right\}
\end{equation} 
where:$\bar R=(R+R_W)$ is the overall Riemann-Weyl curvature scalar, $\beta$ a dimensionless coupling constant, $D_{\rho}$ the Weyl co-covariant derivative ~\cite{foot:Weyl}. According to ~\cite{foot:Weyl} and Eq.~\ref{eq:Beltrami}, in the present context the Weyl curvature scalar may be expressed in the very simple form: $R_W=-6[\phi_\rho\phi^\rho+V'_\Lambda(\phi)]$.
The skew - symmetric tensors are:$\phi_{\rho\sigma}\equiv(\partial_{\sigma}\phi_{\rho}-\partial_{\rho}\phi_{\sigma})$; $f_{\rho\sigma}\equiv(\partial_{\sigma}A_{\rho}-\partial_{\rho}A_{\sigma})$ for the $U(1)_{Y}$ gauge field and: $F^{l}_{\rho\sigma}\equiv(\partial_{\sigma}b^{l}_{\rho}-\partial_{\rho}b^{l}_{\sigma}+g\epsilon_{jkl}b^{j}_{\rho}b^{k}_{\sigma})$ for the $SU(2)_L$ non-abelian gauge fields of the Yang-Mills theory.  Here we introduce the gauge vector bosons: $A_{\rho}$ , i.e. the electromagnetic vector potential, for the group $ U(1)_Y$ and the three components in the isospin space: $\left(b^{1}_{\rho},b^{2}_{\rho},b^{3}_{\rho}\right)$ of the gauge vector boson: $\hat{b_\rho}$ for $SU(2)_L$. Y is the "weak hypercharge" operator. The term proportional to ${\tilde{\lambda}}$ is added for dynamical stability against unbounded field oscillations.  
The dimensionless quantity: $\eta(n)\equiv[(-6+\alpha^{-1})\times n]\simeq(-6n)$, $(\alpha >>1)$  is function of any "real number" (n), either positive or negative, integer or fractional. Note that $\eta(n)=\pm1$ for: $n=\mp(6)^{-1}=\mp0.1667$.
In view of what we believe to be the most interesting result of the present work in what follows we shall deal with the exponential potential functions of $\phi$ and with linear superpositions of these functions. The co-covariance conditions implied by the Weyl symmetry on each addendum $X$ appearing in the expression of $\hat{L}$, i.e. $W(\hat{L})=W(X)=0$, impose well defined constraints on the function $V_\Lambda(\phi)$. Since for any  quantity: $X{\rightarrow{e^{\lambda(x)W(X)}X}}$, and since in the present case: $W(\sqrt{-g})=+4, W(\mu^{2})=-2$, the general co-covariant expression for the potential $V_\Lambda$ with the required value: $W(V_\Lambda)=-2$ may be cast in the form ~\cite{Weyl1952,De Martini2016,francesco2017}: 

\begin{equation}\label{eq:epsilon}
V_\Lambda(n\phi)=C\times{(c^2{\sqrt{G_F}})^{(n-2)}}\times{\exp{[-nK\phi]}}
\end{equation}
where C  is a $\phi$-independent  parameter and (n) is any real number. If the potential is expressed in the alternative form: $V_\Lambda(n\phi)\propto{(c^2{\sqrt{G_F}})^{(n'-2)}}\times{|a|^{- nK\phi}}$, where: $n'\equiv(n\cdot{ln|a|})$, the function: $\eta(n)\equiv(-6n+n\alpha^{-1})\simeq(-6n')$ should appear in Eq.~\ref{eq:LL}. In the case of a linear superposition of exponentials: $\sum_{j}V_\Lambda^{j}(n_j\phi)$ with: $V_\Lambda^{j}(n_j\phi) \propto{(c^2{\sqrt{G_F}})^{(n_j-2)}}\times{\exp{[-n_jK\phi]}}$, or/and : $V_\Lambda^{j}(n_j\phi) \propto{(c^2{\sqrt{G_F}})^{(n'_j-2)}}\times{|a|^{-n_jK\phi}}$, it suffices to plug within the square brackets of Eq.~(\ref{eq:LL}) the corresponding sums:  $\sum_{j}\eta(n_j)V_\Lambda^{j}(n_j\phi)$. Since in the expressions above $(n), (n_j), (a)$ are "real numbers", positive or negative, integer or fractional, the present theory may be defined a "universal theory" ~\cite{Weyl1952,De Martini2016,francesco2017}. In what follows let's  consider the expression given by Eq.~(\ref{eq:epsilon}).
\\
Let us define an "Effective Cosmological Potential":
\begin{equation}\label{eq:efcopo}
V_{eff}(n\phi)\equiv{\left[n\cdot{V_\Lambda(n\phi)} + V'_\Lambda(n\phi)\right]}
\end{equation}
that can be either positive or negative depending on the sign and size of $V_\Lambda(n\phi)$ and/or of its $\phi$-derivative. In case of a negative derivative a very small value of $|V_{eff}|$  may  result  from the sum of two very large contributions with opposite sign. As said, the contribution $V_\Lambda$ is the vacuum energy, i.e. the Dark Energy content of the Universe ~\cite{PenroseHawking1996}. As we shall find in the paper, the quantity $|V_{eff}|\equiv M_U^2 $ contributes to the corresponding \textit{measured} quantity, i.e. to the "cosmological constant", $\Lambda$.   The above effect, expressed by the size of  the "mass-reduction parameter": ${|\zeta}|\equiv\sqrt{{\frac{|V_{eff}|}{|V_\Lambda|}}}<<1$, indeed a general property of the Universe, can lead to a consequence of cosmological relevance since it represents a clue towards the resolution of the "$ \Lambda$-Paradox" ~\cite{Zee2013,Zee2008,Baumann2012}.  A conformally-covariant solution of the Paradox based on Eq. ~(\ref{eq:efcopo}) will be given later in the paper. \\
The exponential potential accounting for the self-interaction of the inflaton field is of large scientific relevance and is also of historical interest: it provided the first clear illustration of an "attractor" solution to the dynamical problem. The idea was championed by: S. Lucchin and S. Matarrese (1985), J. J. Halliwell (1987) and B.Ratra and P.J.E. Peebles (1988). A  rich bibliography on Dark Energy and cosmological inflation are found in the comprehensive review paper by: P.J.E. Peebles and B. Ratra ~\cite{Peebles2002}. Various exponential forms of the $(n\phi)$-potential are also analyzed in a recent book on "Dark Energy" by A. Amendola and S. Tsujikawa ~\cite{Amendola2010}.
Let's consider the condition: $\phi_{\rho\sigma} = 0$\\\\
%%%%%%%%%%%%%%%%%%%%%%%%%%%%%%%%%%%%%%%%%
\section{3.
The Higgs field and the Vacuum Energy in the Universe}
The wide conceptual scenario opened by the preceeding chapter and the structure of the Lagrangian $\hat{L}$, Eq.~(\ref{eq:LL}) offer the possibility of inquiring about the implications of the dark energy content of the Universe within some relevant aspects of the submicroscopic world of the elementary particles. The supposed mass generation properties of the Higgs field, and his pervasiveness that parallels analogous aspects of the inflationary field has already stimulated in recent years a wealth of research in the field referred to as: "Higgs inflation"~\cite{Bezrukov2014,Salvio2015,Hamada2015}.  In what follows we shall find that the connection between the two fields can be demonstrated in the context of the present conformally-covariant theory in which the "classical" GR approach based on the  lagrangian Eq.~(\ref{eq:LL}) is associated with a spontaneously broken $SU(2)_L\otimes U(1)$ gauge theory. We shall briefly consider this theory  in the framework of the standard "\textit{theory of leptons}" due to S. Weinberg, A. Salam, S.L. Glashow ~\cite{Weinberg1964,Salam1964,Glashow1961}. Let us introduce a \textit{complex} iso-doublet of scalar fields ~\cite{Weinberg2013,Georgi2009}: 
\begin{eqnarray}\label{eq:isodoublet}
{\vec{\mu}}\equiv{{\left(\begin{array}{c}\mu^+\\\mu^0\end{array}\right)}}
\end{eqnarray}
that transforms as a $SU(2)_L$ doublet with heavy hypercharge: $Y_\mu = +1$ according to the Gell Mann-Nishijima relation for the electric charge: $Q=I_{3}+\frac{1}{2}Y$. The weak-isospin projection $I_3$ and the weak-hypercharge are commuting operators: $[I_3,Y]=0$ ~\cite{Weinberg2013,Georgi2009}. Next, we introduce in the expression of$\hat{ L}$, Eq.~(\ref{eq:LL}) the following replacement: $D_{\rho}\mu D^{\rho}\mu\rightarrow{(D_\rho\vec{\mu})^\dagger}\cdot{(D^\rho\vec{\mu})}$ where $ D_\rho$ expresses the Weyl-covariant, i.e. co-covariant, derivative, and the dot represents scalar vector multiplication in the isospin space. The gauge co-covariant derivatives are:      
\begin{equation}\label{eq:cocovarderiv}
D_\rho\vec{\mu}= \left[\partial_\rho +  \phi_\rho\right]\vec{\mu}+ \frac{i}{2}\left[g'YA_\rho+ g\hat{\tau}\cdot\hat{ b_\rho}\right]\vec{\mu}
\end{equation}
where the components of the vector $\hat{\tau}$ : $\left(\tau_{1},\tau_{2},\tau_{3}\right) $ are the Pauli spin operators, g represents the coupling constant of the weak-isospin group $SU(2)_L$ and $\frac{g'}{2}$ represents the coupling constant of the weak-hypercharge group $U(1)_Y$. The presence of the inflaton field $\phi_\rho$ multiplied by $\vec{\mu}$ in the first \textit{real} square bracket at the r.h.s. of the Eq.~(\ref{eq:cocovarderiv}) is due to the weight $W(\vec{\mu})=-1$ within the co-covariant derivative~\cite{foot:Weyl}. The Eq.~\ref{eq:cocovarderiv} reproduces the spontaneous symmetry breaking theory of leptons ~\cite{Weinberg2013,Georgi2009} by further introducing within the standard theory the formal change: $\partial_{\rho}\rightarrow{\left[\partial_{\rho}+\phi_{\rho}\right]}$. As shown by  ~\cite{foot:divergence} this change provides a contribution to the interaction between the Higgs and the inflaton fields we are now dealing with. As we shall see by inspection of the lagrangian $\hat{L}$,  Eq.~(\ref{eq:LL}), other larger contributions to the same effect, proportional to $\alpha$, are due to the $\phi_\rho\phi^\rho$ and $ V'_\Lambda(\phi)$ terms in the curvature $R_W$~\cite{foot:Weyl}.
In summary, and most interesting, all that shows that it is precisely the conformally-covariant structure of the Weyl's geometry that establishes the connection between the two universal fields, $\phi$ and $\mu$, the protagonists of the present analysis.\\
Let us briefly outline the standard theory of leptons on the basis of the classic texts ~\cite{Weinberg2013,Georgi2009}. Consider in general terms the kinetic term of the dynamical equation for the scalar field $\vec{\mu}$:  
\begin{equation}\label{eq:kinetic}
\{{(D_{\rho}\vec{\mu})^\dagger}\cdot{(D^{\rho}\vec{\mu})}-{|\lambda|}{\mu^4}\}=\{(\partial_{\rho}\vec{\mu})^\dagger\cdot(\partial^{\rho}\vec{\mu})-{\langle\mu\rangle^2}\mu^2-{|\lambda|}\mu^4\}
\end{equation}
If the mass term ${\langle\mu\rangle}^2$ is negative the continuous symmetry of the system's hamiltonian does not coincide with the symmetry of the vacuum and the condition of "spontaneous symmetry  breaking" takes place. In virtue of a theorem~\cite{Goldstone1962} a possible Goldstone boson is associated with a generator of the gauge group that does not leave the vacuum invariant. We investigate this case by choosing the following vacuum expectation value of the scalar field Eq.~(\ref{eq:isodoublet}):  
\begin{eqnarray}\label{eq:vacisodoublet}
\langle\phi\rangle_0={{\left(\begin{array}{c}0\\\frac{v}{\sqrt2}\end{array}\right)}}
\end{eqnarray}
where, according to the full electroweak theory, the vacuum field is:
\begin{equation}\label{eq:vacuum}
 v=\sqrt{\frac{-\langle\mu\rangle^2}{{|\tilde{\lambda}|}}}
\end{equation}
where the dimensionless parameter: $\tilde{\lambda} =\frac{G_F\cdot{M_H^{2}}c^4}{\sqrt{2}}={0.131}$ , is expressed in terms of: $M_H$, i.e. the mass of the "Higgs boson" we shall define shortly in this Section. The vacuum is left invariant by any group generator G if: $G\langle\phi\rangle_0=0$. In our case we find all the $SU(2)_L$ and $U(1)_Y$ group generators: $\hat{\tau}_{i}, (i=1,2,3)$ and $Y$ operating on $\langle\phi\rangle_0$ break the symmetry of the vacuum. However the $U(1)_{EM}$ symmetry generated by the electric charge preserves the invariance since:
$Q\langle\phi\rangle_0\equiv\frac{1}{2}(\tau_3+Y)\langle\phi\rangle_0=\left(\begin{array}{c}0\\0\end{array}\right)$. The photon, therefore, remains massless and the other three gauge bosons will acquire mass ~\cite{Jammer2000}. These are the heavy bosons: $W_{\rho}^{\pm}\equiv\frac{\left[b_\rho^{1}\mp{i}{b_\rho^{2}}\right]}{\sqrt{2}}$.
and: $Z_{\rho}\equiv\frac{\left[-g'A_\rho+gb_\rho^{3}\right]}{\sqrt{g^2+g'^2}}$. Upon expansion of the Lagrangian~(\ref{eq:kinetic}) about the shifted minimum of the Higgs potential we can investigate the small oscillations around the vacuum $v$ of of the "Higgs field", this one expressed by the field $\theta$.  This dynamics is expressed by the Lagrangian for small oscillations:
\begin{equation}\label{eq:lagrangian2}
\bar{L}={\frac{1}{2}}{\left[(\partial_{\rho}\theta)(\partial^{\rho}\theta)+{2}{\langle\mu\rangle}^{2}\theta^{2}\right]}+\frac{v^2}{8}\left[g^2\mid{b_\rho}^{1}-{i}{b_\rho^{2}}\mid^{2}+(-g'A_\rho+gb_\rho^{3})^2\right]+...
\end{equation}
plus interaction terms. As shown by this equation, the Higgs field has acquired a $(mass)^2\equiv(M_H)^2=2\mid\langle\mu\rangle\mid^2$. The above sentences express in a very summary form the well known results of the standard electroweak theory.\\ \\

We may now inquire about the effects on the theory of the change: $\partial_{\rho}\rightarrow{\left[\partial_{\rho}+\phi_{\rho}\right]}$ introduced in our present analysis by the co-covariant derivative ~(\ref{eq:cocovarderiv}) as well of the general relativistic structure of the overall co-covariant lagrangian $\hat L$~(\ref{eq:LL}).
Note first that  nothing in the standard formulation of the electroweak theory based on the $SU(2)_L\otimes{U(1)_Y}$ group specifies the mass of the Higgs boson and  none of the conceivable applications to the conventional processes depends in any way upon the value of  $M_H$. It may therefore appear that $M_H$ can indeed  be considered a "free parameter" of the standard electroweak theory even if general constraints have been considered by some authors by imposing certain requirements  of internal consistency~\cite{Quigg1983}.\\ 
This is not the case with our present theory. Since the Eq.~(\ref{eq:kinetic}) is formally included into the extended lagrangian $Eq.~(\ref{eq:LL})$, we can easily transfer to the last one all the physical conceptions and the theoretical considerations addressed so far to Eq.~(\ref{eq:kinetic}). In particular, the Riemann curvature $R$ and the Weyl curvature $R_W$ appearing in $Eq.~(\ref{eq:LL})$ are now involved in the spontaneous symmetry breaking scenario of the electroweak theory. On the other hand,  they can also be expressed in terms of the relevant cosmological quantities: i.e. the inflation potential: $V_\Lambda(n\phi)$, its $\phi$-derivative: $V'_\Lambda(n\phi)$ and the inflation vector field:$\phi_\rho$. Consequently, and very important, the dynamical behaviour of all interactions affecting the properties of the elementary particles, including the Higgs field, is directly determined by these cosmological quantities, and then by the overall dynamics of the Universe.  We believe that this is an interesting result. \\

By setting ourselves in a quantum-mechanical perspective we note that, in virtue of the Equations ~(\ref{eq:vacisodoublet}), ~(\ref{eq:vacuum}) the quantity ${v^{2}}\propto{|\langle\mu\rangle|^2}$ can be interpreted as a kind of probability  of generating a massive "Higgs boson" when the energy conditions are met for the generation of this particle out of the vacuum field. Indeed, these energy conditions were attained in the celebrated experiment at CERN in 2012 with the LHC p-p collider operating at the TeV range of energies, i.e. at a collision temperature: $T_c\simeq(10^{15} \div 10^{16}) K$. By that experiment the mass of the Higgs boson was measured: $M_H={125.09}\pm{0.21} (GeV/c^2)$ ~\cite{Aad2012,Chatrchyan2012,Tonelli2015}.\\
According to the standard electroweak theory, the vacuum energy-density in the Universe is expressed in general by (~\cite{Veltman1975,Peebles2002,Quigg1983}):
\begin{equation}\label{eq:rhovac}
 \rho_{\Lambda}=\frac{(M_H^{2}c^4\cdot v^2)}{8}
\end{equation}
In the case of the "artificial" generation of the Higgs boson by the CERN experiment we should set: $(v^2) =(\sqrt{2}G_F)^{-1}\simeq{6.07}\cdot{10^{4}}  (GeV)^{2}$  leading to the size of the contribution of the Higgs field to the vacuum energy density: $\rho_{M_H}\simeq(1.18\cdot10^8) (GeV)^4 $.  While it is conceivable that the suitable energy conditions for the "spontaneous" generation of the Higgs bosons were actually realized in the very early Universe, at the present time, characterized by a CMB temperature as low as $T_0=2.725 K°$, i.e. $K_BT_0=2.348\cdot10^{-4} (eV)$, the quantity:$(v^2)=\frac{|\langle\mu\rangle|^2}{\tilde{\lambda}}$ could be interpreted as to express an - extremely small - $(mass)^2$ probability-density associated with the present size of the vacuum field in the Universe. In support of this argument note that the ratio of the energy densities $(\rho_{M_H}/{\rho_\Lambda})$, where $\rho_\Lambda$ is of the order of ${(1 meV)}^4$ - as reported below in Sect. 5 - scales approximately as: $\delta^{4}$, being: $\delta=\frac{(K_B{T_c})}{(K_B{T_0})}\sim{10^{15}}$, as expected. This probabilistic interpretation will be resumed in the following Sections, where the $\Lambda$-paradox will be considered explicitly.\\\\

We may analyze the symmetry-breaking condition and evaluate the quantity ${|\langle\mu\rangle|^2}$ by variation of the lagrangian Eq. ~(\ref{eq:LL})  respect to the metric tensor $g_{\rho\sigma}$. This leads to a somewhat clumsy expression including several  multiple derivatives of the fields $\mu$ and $\mu^2$. The procedure is simplified by making appeal to the conformal invariance of the theory by choosing the gauge: $\mu(x)=\mu_o$, constant. In this gauge the metric tensor and the field $\phi$ are related to the same quantities adopted in the previous gauge by the conformal transformations: $g'_{\rho\sigma}=(\mu(x)/\mu_o)g_{\rho\sigma}$ and:$\phi'(x)=\phi(x) + ln(\mu(x)/\mu_o)$. We adopt here the simplifying approximation $(\mu(x)/\mu_o)=1$ which expresses the constancy of the size of the mass of the elementary particles in the Universe. Accordingly, we keep the same symbols of the former equations, i.e.:  $g'_{\rho\sigma}=g_{\rho\sigma}$,  $\phi'(x)=\phi(x)$, $R'=R$. Within this approximation, the Euler-Lagrange equation consists of the Einstein equation which may be cast in the form:
\begin{equation}\label{eq:Einstein2}
 R_{\rho\sigma}-\frac{1}{2}g_{\rho\sigma}R +  g_{\rho\sigma}{\bar{\eta}}\left [V_{eff}(n\phi)+\frac{1}{2}{\phi_\lambda\phi}^\lambda\right ] = K^2 T_{\rho\sigma}
\end{equation}
where:$ \bar{\eta}\equiv[3+{(2\alpha)}^{-1}]\simeq 3$. We also include the energy-momentum tensor $T_{\rho\sigma}$ due to the gauge fields appearing in Eq.~(\ref{eq:LL}).   This equation is  interesting because it shows that the generally very large "Inflation Potential" $|V_{\Lambda}(\phi)|$ appearing in the Lagrangian $\hat{L}$, Eq.~(\ref{eq:LL}) is replaced here by the far smaller: $|V_{eff}(n\phi)|$. An approximate expression of the  Riemann curvature is found by taking the trace of both sides of the above equation:
\begin{equation}\label{eq:R}
R_\Lambda=\bar{\eta}\left[2V_{eff}+\phi_{\lambda}\phi^{\lambda}\right]
\end{equation}
where the effect of the Yang-Mills fields is neglected in this approximetion since it contributes with a term proportional to: $(\alpha)^{-1}<<1$. The Eq.~(\ref{eq:R}) expresses the value of the Riemann curvature in an empty Universe filled with the vacuum- and  inflaton- fields. The vector field $\phi^\lambda$ will be considered is Section 6 as a possible candidate for Cold Dark Matter (CDM). The substitution: $R_\Lambda \rightarrow  R$ within Eq.~(\ref{eq:LL}) leads immediately to the spontaneous symmetry-breaking term:
\begin{equation}\label{eq:Hi}
(\langle\mu\rangle)^2=  6\alpha{V_{eff}}
\end{equation}
and, in virtue of Eq.~(\ref{eq:kinetic}) to the spontaneous symmetry-breaking condition: $V_{eff}<0 ,  i.e.:  V_{eff}=-|V_{eff}|$.\\ 

The structure of the above equations may suggest an interesting interpretation. Note that Eqs. ~(\ref{eq:Einstein2}) and ~(\ref{eq:R}) may be expressed in the form:  
\begin{equation}\label{eq:Einstein3}
 R_{\rho\sigma}-\frac{1}{2}g_{\rho\sigma}R + g_{\rho\sigma}\tilde{\Lambda} = K^2 T_{\rho\sigma}
\end{equation}
where the last term in the r.h.s., here proportional to: $\tilde{\Lambda}=\frac{1}{2}(R_\Lambda)>0$, was added "by hand" by Einstein to his famous equation. It may appear interesting that, in virtue of  the curvature $R_W$, the Weyl geometry leads  to a structure close to the "complete" Einstein equation with no artificial manipulations.\\
 The "measured" value of the "cosmological constant", $\Lambda$ will be evaluated in Section 5 on the basis of the PLANCK experimental data. In the same Section we shall find the theoretical result: $\Lambda=6|V_{eff}|$ by the adoption of the full electroweak theory. The highly significant  identification: $\Lambda = \tilde\Lambda$ would impliy to the following interesting expression for the "norm" of the Weyl vector field: $\phi_{\lambda}\phi^{\lambda} = \Lambda$.\\

%%%%%%%%%%%%%%%%%%%%%%%%%%%%%%%%%%%%%%%%%%%%%%%%%%%%%%%%%%
\section{4.
The Cosmological Constant Paradox and its solution}

The following quote by Richard Feynman (1965) is enlightening~\cite{Feynman1965}: \textit{"..Such a mass density would, at first sight at least, be expected to produce very large gravitational effects which are not observed. It is possible that  we are calculating in a naive manner, and, if all of the consequences of the general theory of relativity (such as the gravitational effects produced by the large stresses implied here) were included, the effects might cancel out; but nobody has worked all this out. It is possible that some cutoff procedure that not only yields a finite energy density for the vacuum-state but also provides relativistic invariance may be found. The implications of such a result are at present completely unknown."}.\\\\

Let us analyze the $ \Lambda$-paradox  in \textit{"a naive manner"}, as in Feynman's words. The vacuum energy density in the Universe $\rho_{vac}$, i.e. the "zero-point energy" of the quantum fields associated with all existing quantum particles is evaluated by (QFT). For simplicity we consider here only the "photon", which is the subject of (QED) (Quantum Electro-Dynamics), the chapter of (QFT) accounting for the electromagnetic (e.m.) phenomena. To carry out the calculation of $\rho_{vac}$ we should first evaluate the spatial density of the available $\vec{ k}-modes$, i.e. of the spatial vectors over which the photons propagate in the free space. Afterwards, each mode, which is modelled by QED as a quantum-mechanical oscillator, is multiplied by 2-times (because of the two orthogonal polarizations) the oscillator's "zero-point energy", which is $(h\omega/2)$, where $(h)$ is the Planck constant. The frequency $\omega$ could be taken to range from zero to a limit cutoff that is determined by the shortest length considered in Physics, i.e. the Planck length, $l_{P}=1.616\times{10^{-33}}(cm)$: ${\omega_{c}}=({2\pi{c}}/{l_{P}})$. The size of final vacuum-energy density is found:   $\rho_{vac}=\frac{2\pi\hbar{c}}{L_{P}^{4}}=\frac{2\pi(c^7)}{\hbar(G_N)^2}={2.9}\times{10^{98}}{(Joules/cm^3)}={1.8}\times{10^{108}}{(GeV/cm^3)}$. To be more realistic the frequency cutoff  $\omega_{c}$ could be determined by the size of the Compton wavelength of the proton: $\lambda_c\simeq 2\cdot10^{-14}$ cm. In this case the energy density given above should be multiplied by the factor $\sim10^{-76}$ leading to: $\rho_{vac}\simeq{1.8}\times{10^{32}}{(GeV/cm^3)}={(1.15\times {10^8})^4} (GeV)^4 $.  Even in this case the size of the numbers keeps being impressive.   We should remind here that  in the vast realm of modern Science, the (QED) theory and, in particular, the "vacuum-field" concept were the most tested paradigms, ever. Uncountable and impressively precise experiments involving the atomic physics, the optical spontaneous emission, the Casimir effect, the atomic spectra, the Lamb's shift, the electron's magnetic moment etc. were the landmarks of the great success of the $ XX$ century Physics.  The calculation carried out for photons should now be extended to the other existing  particles and the corresponding energy contributions should be added, without any reasonable chance of mutual cancellations.\\ 
The paradox consists of the mysterious, humongous discrepancy existing between the enormous size of the overall calculated: $ V_\Lambda$ and the very small size of the "Cosmological Constant" $\Lambda\propto{V_{eff}}$, measured today.\\ 
 An authoritative early review on the cosmological constant problem is due to S. Weinberg ~\cite{Weinberg1989}. In more reent years comprehensive reviews on the same problem have been contributed by: R. Bousso and J. Polchinski (2000), P.J.E. Peebles and B. Ratra (2002),  L. Dyson, M. Kleban and L. Susskind (2002), J. Polchinski (2006), R. Bousso and B. Freivogel (2006), P. J. Steinhardt and N. Turok (2006), A. Zee (2008 and 2013)  ~\cite{Peebles2002,Zee2008,Zee2013,Polchinski2006}.\\ \\ 
Our present  theory offers a conformally-covariant solution of the $\Lambda$-paradox. Consider the following co-covariant potential derived by Eq. ~(\ref{eq:epsilon}) which accounts for the overall vacuum-energy content of the Universe calculated by the methods of (QFT) shown above:

\begin{equation}\label{eq:EPS}
V_\Lambda(\phi)=C\times{(c^2{\sqrt{G_F}})^{(n-2)}}\times{\exp{[n(\epsilon-1)]K\phi}}
\end{equation}
The quantity $\epsilon$ is a dimensionless real number, positive or negative: $\epsilon\approx0$. Plugging Eq.~(\ref{eq:EPS}) into Eq.~(\ref{eq:efcopo}) one is led to the ratio: $\frac{{V}_{eff}}{{V}_\Lambda}=n\epsilon$. This is indeed a most drastic "mass-reduction" effect with the parameter: $|\zeta|=\sqrt{|n\epsilon|}<<1 $  ~\cite{Baumann2012}. By this argument, the enormous content of vacuum-energy in the Universe, calculated by (QFT) and  expressed by $|{V}_\Lambda(n\phi)|$  is made consistent by our theory, i.e. by our Eqs.~(\ref{eq:epsilon}) and ~(\ref{eq:efcopo}), with the very small value of: $|V_{eff}(n\phi)|=|{n\epsilon}\cdot{V_\Lambda(n\phi)}|$  and then, of the "effective" cosmological constant measured today ~\cite{Dodelson2003,Weinberg1972}. Therefore the $\Lambda$-Paradox is resolved in D=4, for any (n).\\
A semi-classical resolution of the paradox cannot be obtained in the context of the D=4 Riemann's geometry.\\ 
%%%%%%%%%%%%%%%%%%%%%%%%%%%%%%%%%%%
\section{5.
Experimental results}

 Let's take ${L_{universe}}$ to express the linear dimension of the Universe. The mass $M_U\propto{\frac{1}{L_{universe}}}$ has been defined by Antony Zee as a sort of "Compton mass of the Universe" and is proportional to the Hubble radius $H_o$ ~\cite{Dodelson2003,Zee2013}. Accordingly, $M_U$  is a quantity accessible to the measurements by the modern space experiments, e.g. by the recent Mission PLANCK. This last experiment gives a rather precise value of the curvature contributions to the present cosmological model: ${\Omega_K}\approx{0}$  ~\cite{PLANCK}.\\
The Equation 21 of the PLANCK paper reports the experimental datum: ${\Omega_m}=0.308\pm{0.012}$ from which, with Equation 50: ${\Omega_\Lambda}=0.692\pm{0.017}$. This represents the curvature contributed by the "cosmological constant". Furthermore, the PLANCK paper reports the cooperation value: ${(67.6\pm{0.6})} (km\cdot{s^{-1}}\cdot{Mpc^{-1}})$ corresponding to the "empirical" value of the Hubble constant: ${H_o}={(2.191\pm{0.02})}\cdot{10^{-18}} (s^{-1})$.\\

These data imply an "empirical" value of the cosmological constant:  ${\Lambda}=3{\Omega_\Lambda}{H_o}^2=99.6\cdot10^{-35}s^{-2}=3.86\cdot10^{-64} (eV/c^2)^2$. The corresponding energy-mass value is : $M_{\Lambda}c^2 =\sqrt{\Lambda}\hbar\approx(9.746\cdot10^{-34})$ (eV). The value of energy density related to $\Lambda$ can be taken to be proportional to the square of the geometric mean, i.e. $({M_\Lambda}\times{M_P})\equiv{\bar{M}^2}$, of the energy-mass $M_\Lambda$ and of the Planck mass: $M_{P}= 1.22\times{10}^{19} (GeV/c^2)$ ~\cite{Zee2013}.  As a result, the experimental value of the energy density is: $\rho_{\Lambda}\approx(3.44\cdot10^{-3})^4 (eV)^4 =1.4\cdot10^{-44} (GeV)^4$.\\

In order to make connection with our theory, it is interesting to consider the analogous results evaluated in the scenario of the full electroweak theory ~\cite{Peebles2002,Veltman1975,Quigg1983}. In that framework the vacuum energy density $(\rho_{\Lambda})$ is given by Eq.~(\ref{eq:rhovac}) and the cosmological constant is: $\Lambda=(\rho_{\Lambda})\times(\frac{8\pi(G_N)}{\hbar(c^5)})$. We may now express the vacuum field $(v^2)$ in terms of the field ${|\langle\mu\rangle|}^2$ which is proportional to: ${{M_U}^2}$ in virtue of Eqs. (~\ref{eq:vacuum}), and (~\ref{eq:Hi}). By further inserting in Eq.(~\ref{eq:rhovac}) the given expressions for $(\alpha)$ and $(\tilde{\lambda})$ it is finally found: $\Lambda=6|V_{eff}|$. This result shows that the present value of the "mysterious" \textit{Cosmological constant} of the Universe is determined by a mere algrebraic "mismatch" of the two counteracting exponential potential functions of the "inflaton" field entering in the expression of $V_{eff}(n\phi)$,  Eq.(~\ref{eq:efcopo}). In other words, the $\Lambda$-paradox would not exist, i.e. $\Lambda =0$, if the mathematical structure of $V_{eff}$, Eq.~(\ref{eq:efcopo}), would be realized exactly, i. e. including the exponentiality of $V_\Lambda$  and the correct linear function: $\eta(n).$ \\  This result is important and expresses the key motivation of the present work.\\\\
Here come the bad news !  Note that the small value of $|V_{eff}|$  results from the algebraic sum of two very large numbers representing the enormous size of: $V_\Lambda$ and of:$V'_\Lambda$. This makes extremely unstable the dynamical system of the Universe. If under the effect of any dramatic catastrophic event (e.g. a collision of gigantic black-holes, or the quantum fluctuation of some relevant cosmological parameters) the value of $V_{eff}$ becomes zero or changes sign, the drama will turn into tragedy, since the vacuum system suddenly symmetrizes. Then, the Higgs field and all masses disappear and the Universe (including ourselves) will  be blown up in an enormous flash of radiation. This fearful instability condition has been analyzed within the framework of the electroweak theory, among others, by: ~\cite{Degrassi2013}.\\ \\

%%%%%%%%%%%%%%%%%%%%%%%%%%%
\section{6.
A meson vector field: Dark Matter ?}
If the "integrability condition" of the Weyl geometry, i.e.  $\phi_\rho=\partial_\rho\phi$ and:  $\phi_{\rho\sigma}=0$, is relaxed, the quantities $\phi$ and $\phi_\rho$ must be considered independent variables of the theory and this one gets even richer ~\cite{foot:VV}. In particular, the inner structure of the  Weyl geometry does not change, since the parallel displacement of vectors involves directly the vector-field $\phi^\rho$, as said. Leaving aside all complications and deferring an exact analysis to future work, we may immediately apply the mathematical methods adopted in the previous Sections to the "geometrical" part of the Lagrangian $\hat{L}$,  Eq.~(\ref{eq:LL}). The variation respect to the \qo{\textit{Weyl vector}} $\phi_\rho(x)$  leads, via the gauge-fixing $\partial_\rho\phi^\rho=0$, to the following Proca equation expressing the dynamics of a massive vector-meson, $\phi^\rho$: 
\begin{equation}\label{eq:Proca}
\left[\nabla_B+(\xi\frac{M_{P}c}{\hbar})^{2}\right]\phi^\rho=0
\end{equation}
where $\xi=\sqrt{2}{(\beta)^{-1}}$ expresses the coupling of the particle to gravitons. It has been found that the massive meson $\phi^\rho$, while strongly interacting with gravitons, does not interact with any $spin-\frac{1}{2}$ or $spin-1$ elementary particle of the Standard Model. In other words, it is not coupled minimally to photons, hadrons and light or massive leptons~\cite{HochbergPlunien1991,Chen1988}. All these properties make this geometrical entity eligible for being considered an optimum candidate for (CDM), the elusive object wich is now under investigation in a large number of laboratories around the world. As it is well known, the (CDM) may consist of a Weakly Interacting Massive Particle (WIMP), a stable SUSY particle, a light neutralino with a mass of the order of $10^2 ( GeV/c^2) $ or even an "axion" particle with a mass as low as $10^{-5}(GeV/c^2)$ ~\cite{Sanders2016,Amendola2010}. Large concentrations of (CDM) have been detected in zones of the Universe characterized by a large inhomogeneous gravitational energy density. Indeed the CDM predominately clusters on the scale of galaxies. In summary, today the size of the mass and any other physical property of (CDM) is inferred from some sophisticated quantum theory based on an attributed "physical model".\\
The field norm $(\phi_\rho\phi^\rho)$ appearing in the Eqs.~(\ref{eq:R}),~(\ref{eq:Einstein3}) may be contributed by the meson-vector field analyzed in the present Section.  However, a convincing experimental evidence of the existence of the vector field $\phiì^{\lambda}$ is today missing and the dynamics of that norm is unknown. \\\\
%%%%%%%%%%%%%%%%%
\section{7.
Conclusions}
The present work shows that the result of the experiments carried out on the cosmic vacuum-energy, or Dark Energy  by our measurement apparata consists of the reduced value of the effective cosmological potential, $V_{eff}$ and not of the full value of $V_{\Lambda}$. At the root of this effect lies the most basic aspect of the Weyl geometry, i.e. the "calibration" (or "gauge") process within the parallel displacement of vectors in curved spaces. This mathematical process proves to be a fundamental, intrinsic feature of all measurement processes in spacetime and reveals itself when the experiments are carried out over cosmological spaces and times. The geometrical mechanism proposed by the present work appears to represent a unifying scenario in which the Weyl scalar curvature $R_W$ plays a surprisingly large role in determining the evolution of the Universe \qo{at large} as well as, at the microscopic level, of the everyday quantum phenomenology~\cite{foot:QCD,De Martini2016}. We believe that this very general dynamical effectiveness is a consequence of the \textit{local} character of the scale-invariance gauge implied by the Weyl's geometry.  All that may appear to be a glimpse into quantum gravity. \\
Thanks are due to Enrico Santamato for valuable advices and collaboration.\\

%%%%%%%%%
%\bibliography{qmzshort,footnotes}

\begin{thebibliography}{24}
\bibitem[{foot({\natexlab{a}})}]{foot:divergence}
\bibinfo{note}{In the context of $\hat{L}$ we can set the value of the covariant divergence: $\frac{1}{\sqrt{-g}}\partial_{\rho}[\sqrt{-g}\phi^{\rho}{(\vec{\mu}^\dagger}\cdot{\vec{\mu}})}]=0. $
Then:
$\phi^\rho_{\mid\rho}{(\vec{\mu}^\dagger}\cdot{\vec{\mu}})=-\phi^{\rho}\partial_{\rho}{(\vec{\mu}^\dagger}\cdot{\vec{\mu}})$
and: ${\left[(\partial_{\rho}+\phi_{\rho})\vec{\mu}^\dagger\right]\cdot\left[(\partial^{\rho}+\phi^{\rho})\vec{\mu}\right]}=\{(\partial_{\rho}\vec{\mu})^\dagger\cdot(\partial^{\rho}\vec{\mu})+\left[\phi_{\rho}\phi^{\rho}+V'_\Lambda(\phi)\right]{\mu^2}\},\:
$in virtue of Eq~(\ref{eq:Beltrami}).  As we see by inspection of the lagrangian $\hat{L}$,  Eq.~(\ref{eq:LL}) the most important contributions to the vacuum symmetry breaking are due to the term: $(\alpha\mu^2\cdot{\bar{ R}})$.
\bibitem[{foot({\natexlab{a}})}]{foot:VV}
\bibinfo{note}{In particular, the cosmolgical potential should now be dependent both on $\phi$ and $\phi^\rho$,i.e.: $V_\Lambda(\phi,\phi^\rho)$.}
\end{thebibliography}
%%% Note: reference Montani et al. was modified by hand (Santamato)

%%%
\end{document}